# KONUS Beam Dynamics Design of Uranium IH-DTL for HIAF

Dou Wei-Ping(窦为平)[a]  He Yuan(何源)[a]  Lu Yuan-Rong(陆元荣)[b, 2]

[a] Institute of Modern Physics, China Academy of Science, Lanzhou 730000, China

[b] State Key Lab of Nuclear Physics and Technology, Peking University, Beijing 100847, China

**Abstract:** KONUS beam dynamics design of uranium DTL with LORASR code is presented. The $^{238}U^{34+}$ beam, whose current is 5.0 emA, is accelerated from injection energy of 0.35 MeV/u to output energy of 1.30 MeV/u by IH-DTL operated at 81.25 MHz in HIAF project at IMP of CAS. It achieves the transmission efficiency of 94.95% with the cavity length of 267.8 cm. Optimization aims are the reduction of emittance growth, of beam loss and of project costs. Because of the requirements of CW mode operation，the designed average acceleration gradient is about 2.48MV/m. Maximum axial field is 10.2MV/m, meanwhile Kilpatrick breakdown field is 10.56MV/m at 81.25MHz.

## 1 Introduction

HIAF is high intensity accelerator facility at IMP of CAS. HISCL is high intensive heavy ion superconducting linear accelerator for injector of HIAF. It consists of ion source, LEBT, RFQ, MEBT, superconducting linear accelerator. It provides proton to uranium beam of 25MeV/u for synchrotron. Up to now, RFQ beam dynamics has been designed with DESRFQ [1] code, which is developed by ITEP (Institute for Theoretical and Experimental Physics), especially for the RFQ with an external buncher. Several QWRs were planned to accelerate the beam from the energy of 0.35MeV/u to 1.3MeV/u. Because of the high shunt impedance of IH-DTL [2][3] for the low β particles, a normal conducting IH-DTL cavity based on KONUS [4][5][6] beam dynamics is proposed to replace the original QWRs to shorten the length of accelerators and decrease the manufacturing cost.

KONUS (KOmbinierte NUll grad Struktur–KONUS）beam dynamics，which means “Combined Zero Degree Structure”，can well overcome confliction of transverse defocusing, longitudinal bunching and accelerating of RF fields. Many projects, such as GSI High Charge State Injector [7], High Current Injector [8][9], and Heidelberg Therapy Injector [10]，have been designed, manufactured and operated successfully by using KONUS beam dynamics. They show IH-DTL for low β ions has very high accelerating gradient.

[1]Supported by NSFC, Grand No.11079001

[2]Corresponding author: yrlu@pku.edu.cn

A KONUS period consists of a quadruple triplet， a rebuncher section with traditional negative synchronous phase and a multi cells acceleration section with zero degree synchronous phase. KONUS beam dynamics simulation is coded to LORASR， which is abbreviated from German “LOngitudinale und RAdiale Strahldynamikrechnungen mit Raumladung”[11]. In normal linear accelerator, the synchronous particles with 0° will have the maximum kinetic energy gain, but the stable phase range becomes zero. When bunch center phase is injected at a bit positive, radial motion is focused and longitudinal motion is defocused. Bunched particles have less energy gain than zero degree synchronous phase particle. If they have a bit higher injection energy, they will arrive at the accelerating gap slight earlier than RF ramping, i.e. bunched center phase moves from positive to negative phase gradually. So Radial motion moves from focusing to defocusing，meanwhile longitudinal motion from defocusing to focusing. More zero degree synchronous phases are setting until accumulated radial defocusing needs to be compensated by the quadruple lenses. After quadruple lenses， because of the beam energy spread，longitudinal needs to be rebunched at the beginning of each KONUS section.

## 2 The IH-DTL Simulation Results of LORASR

Input twiss parameters for LORASR are listed in Table 1. Phase advance per structure period is shown in Figure 1. Input and exit particles distribution are shown in Figure 2. The design of RFQ is under optimization and MEBT is used to match beam parameters from exit of RFQ to entrance of IH-DTL. Converging beam with big size and small angle is needed for the better beam transmission.

Table 1. Input tiwss parameter

| Input twiss parameters | $\alpha$ | $\beta$ (mm/mrad) | $\varepsilon_{n.rms}$ (mm.mrad) |
|---|---|---|---|
| x | 0.9064 | 0.5567 | 0.2041 |
| y | 0.9636 | 0.5005 | 0.2024 |
| z | 0.3188 | 0.3989 | 0.1228 |

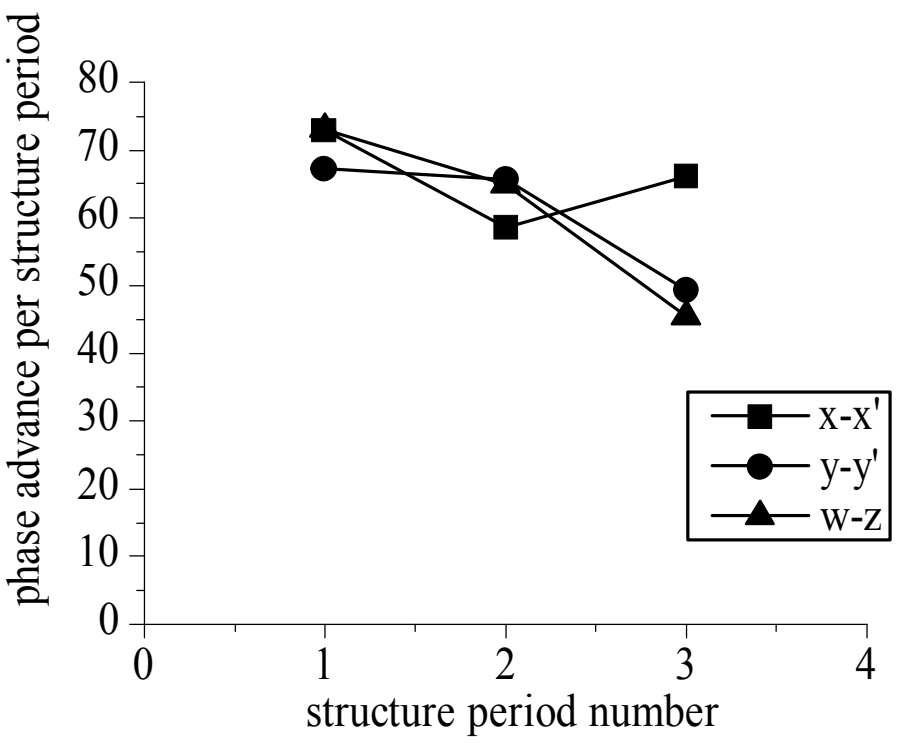

Fig 1 Phase advance per structure period

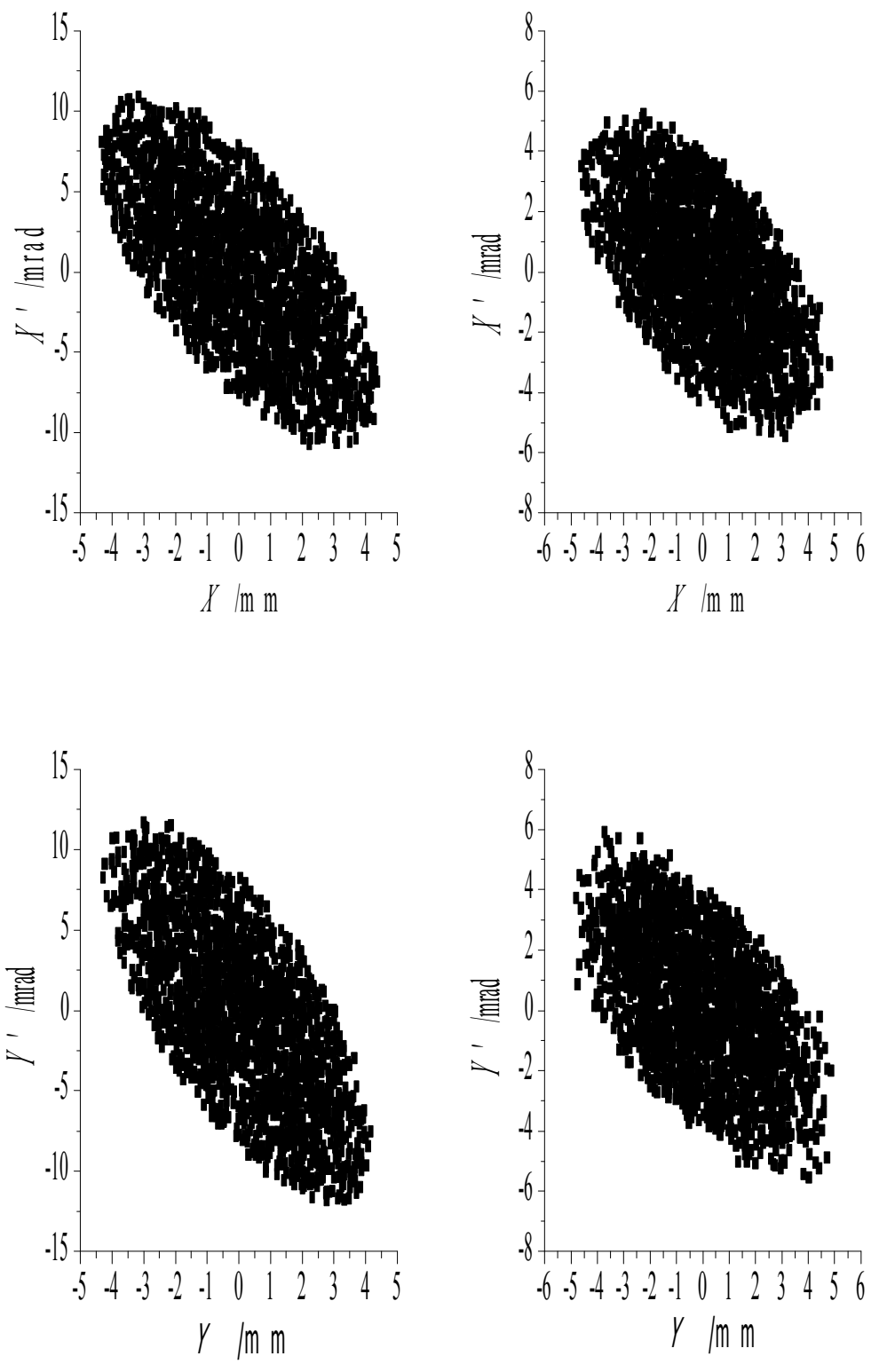

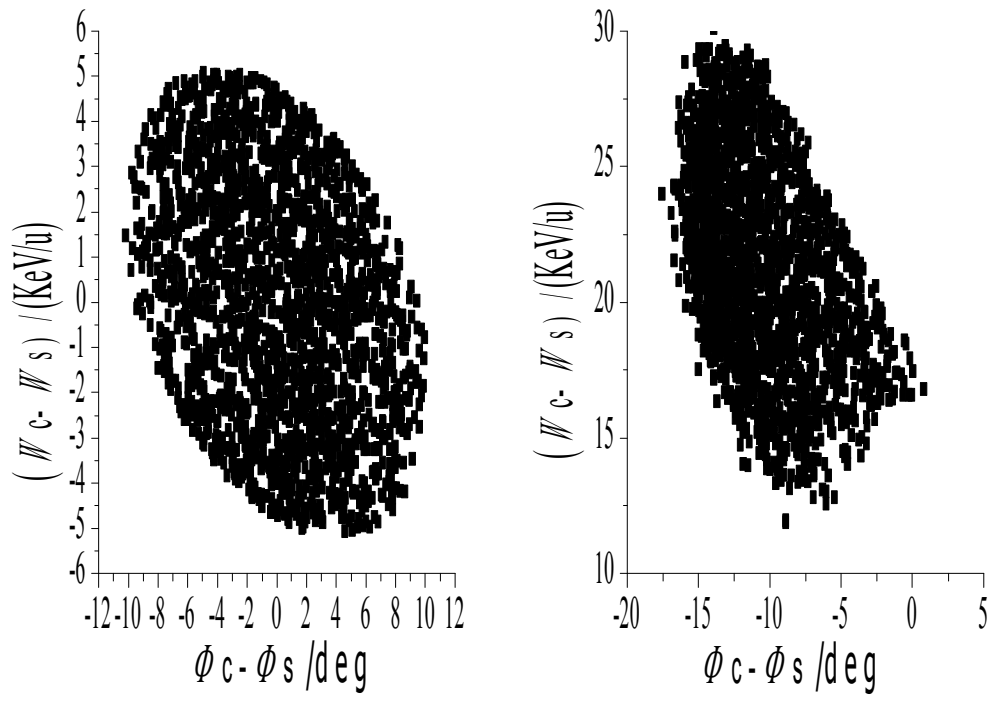

Fig 2 Left: Input particles distribution (100%) Right: exit particles distribution (100%)

Energy spread and synchronous phase at every accelerating gap for different KONUS sections (for any particle index as “c”, synchronous particle index as “s”) is shown in Figure 3. Conventional negative synchronous phase sections are needed in front of zero degree sections (Figure 3 , position ‘a’) to keep longitudinal motion be focused. when the particles pass first gap of zero degree sections (Figure 3, position ‘b’, ‘d’ and ‘f’), the bunched center phase is at a bit positive. Radial motion is focused and longitudinal motion is defocused. But bunched center particles have a bit higher injection energy. They will arrive at the accelerating gap slight earlier than RF ramping, i.e. bunched center phase moves from positive to negative phase gradually. At exit (Figure 3 , position ‘c’, ‘e’ and ‘g’), Longitudinal motion is focused and radial motion is defocused. The longitudinal (Figure 4a and 4b) and transverse (Figure 4c and 4d) beam 100% envelopes for the design current of 5.0emA are shown in Figure 4. Figure 4.a and 4.b shows that the bunched beam center has the same kinetic synchronous energy and synchronous phase with the designed one inside three rebuncher drifting gap sections. However the kinetic beam energy at bunched beam center is a bit higher than that at any gap center for all KONUS zero degree drifting tube sections and triplet lenses, which brings a matched phase slip against the zero degree synchronous particles. The dash line in Fig 4.b shows the phase distribution of bunched beam center particles. Figure 4.c and 4.d shows the transverse beam envelopes are less than 4.6–6.0 mm within the resonators and up to 7.2 mm within the lenses. The beam aperture diameter of the triplets is 24.0mm and the drift tube sections with inner aperture diameters of 18.0-20.0mm. Zero degree synchronous phase section in IH-DTL compared to traditional negative phase kept in Alvarez structure brings weaker radial defocusing effect and higher longitudinal energy gain. In addition ， bunching in negative synchronous phase and focusing in triplet, make bunched particles stable in three planes and bring less emittance growth. Figure 5 gives the results of normalized emittance growth, which is about 8.0%. Figure 6 represents the transmission efficiency as a function of input beam current.

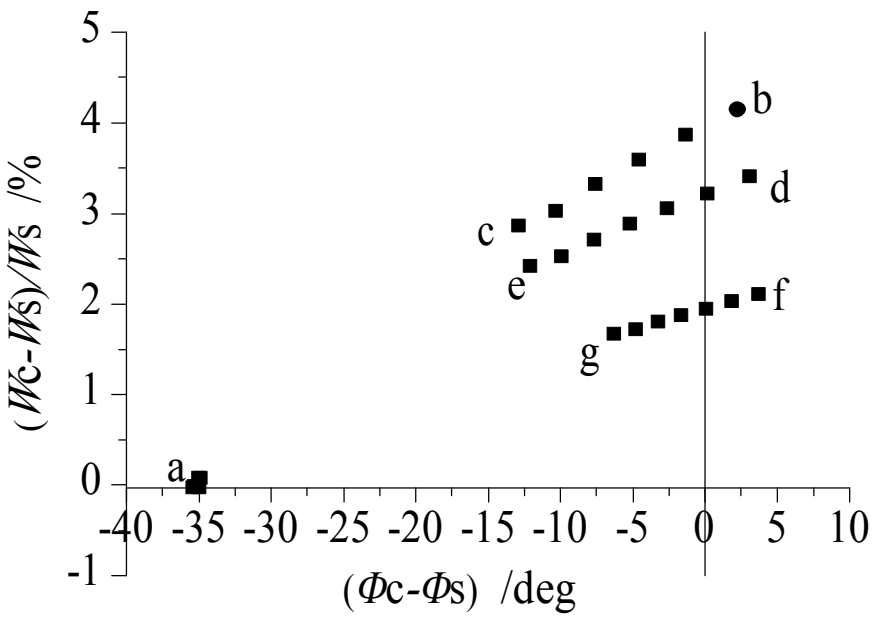

Fig 3 Energy spread and synchronous phase at every accelerating gap（from ‘a’ to ‘g’）

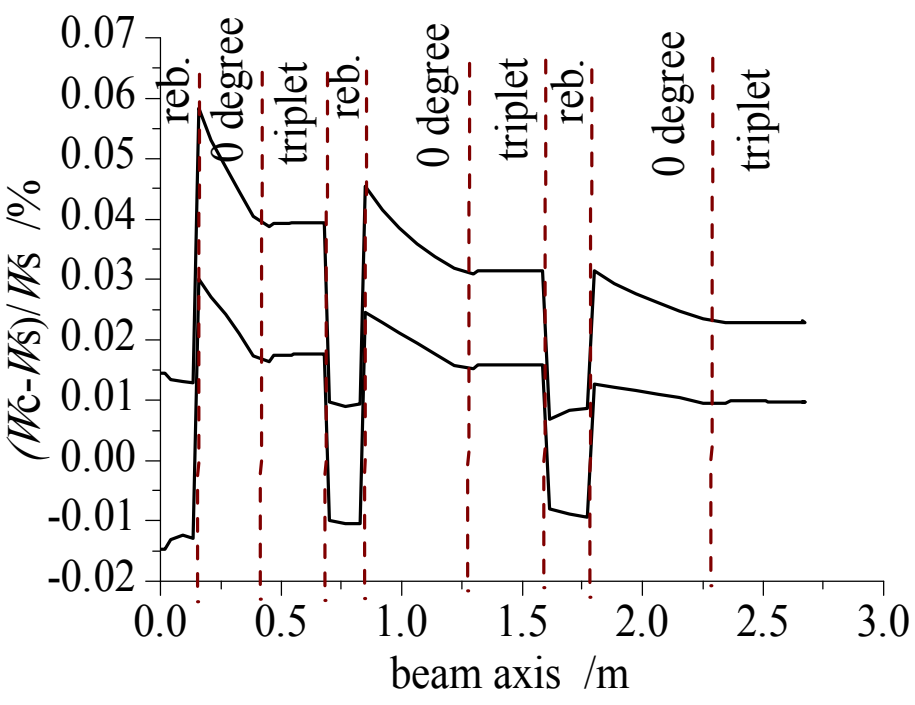

Fig 4.a 100% Envelopes in longitudinal direction

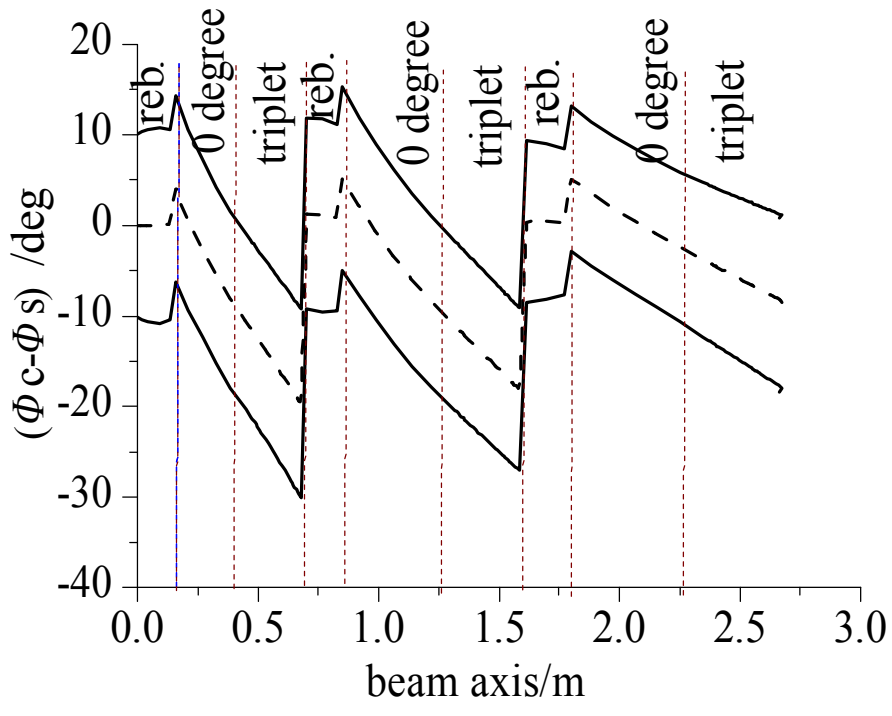

Fig 4.b 100% Envelopes in longitudinal direction

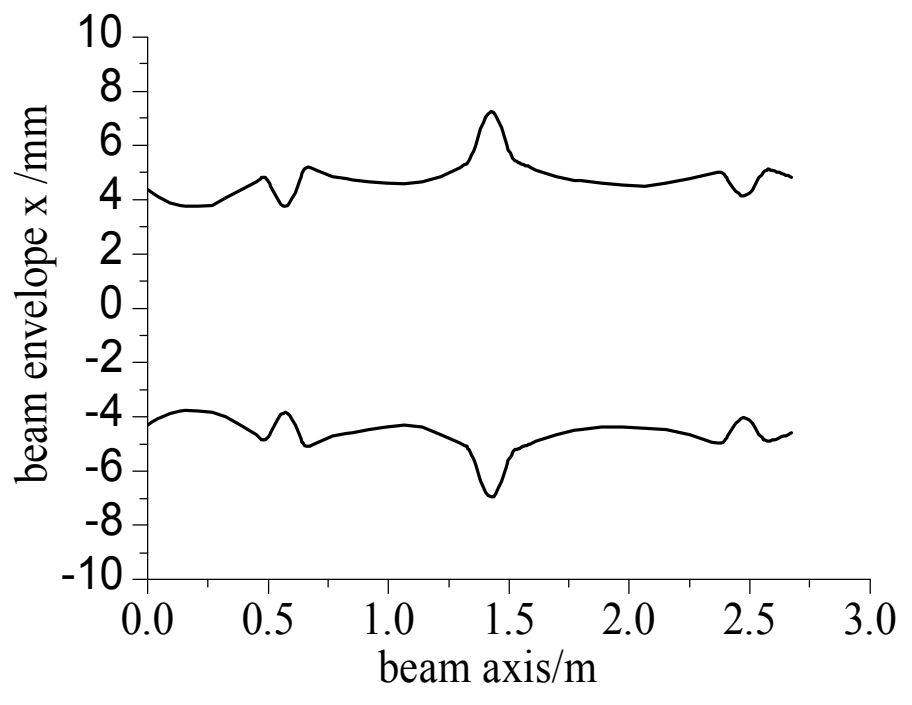

Fig 4.c 100% Envelopes in xz direction

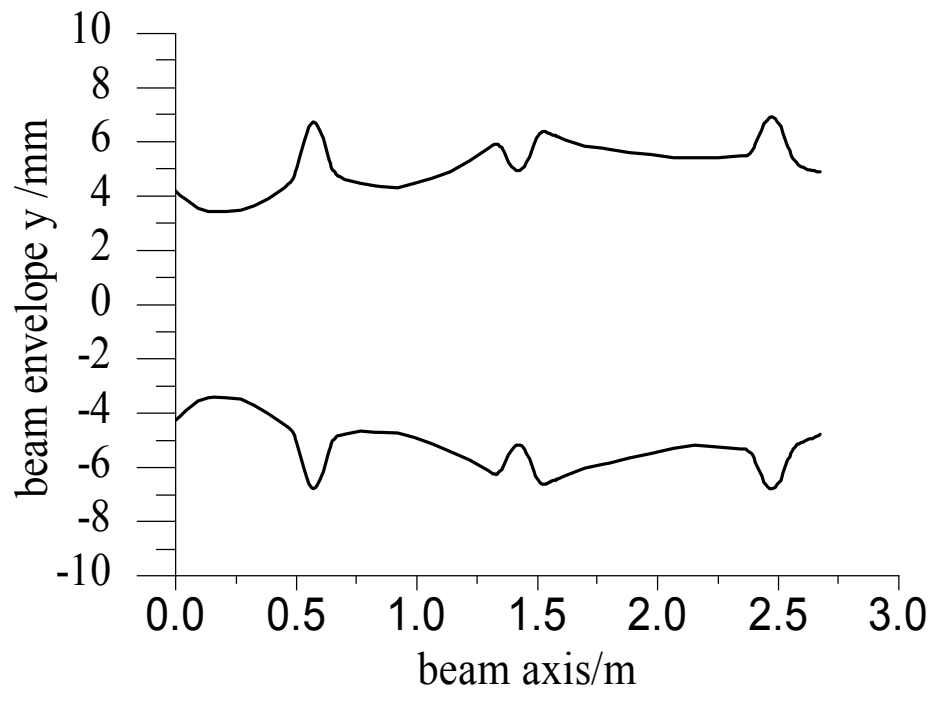

Fig 4.d 100% Envelopes in yz direction

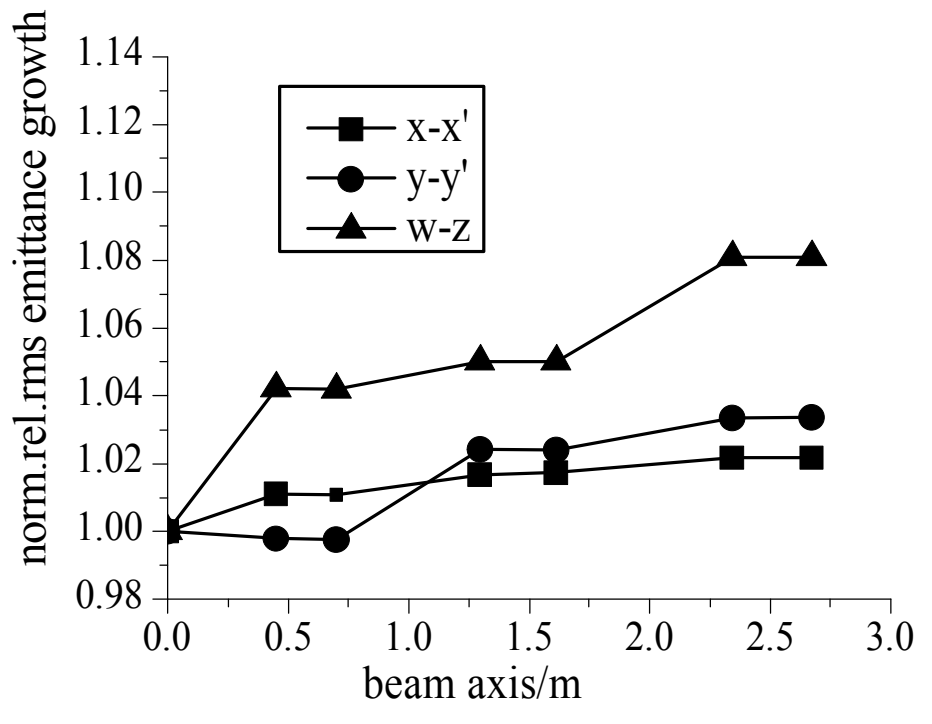

Fig 5 Normalized RMS emittance growth

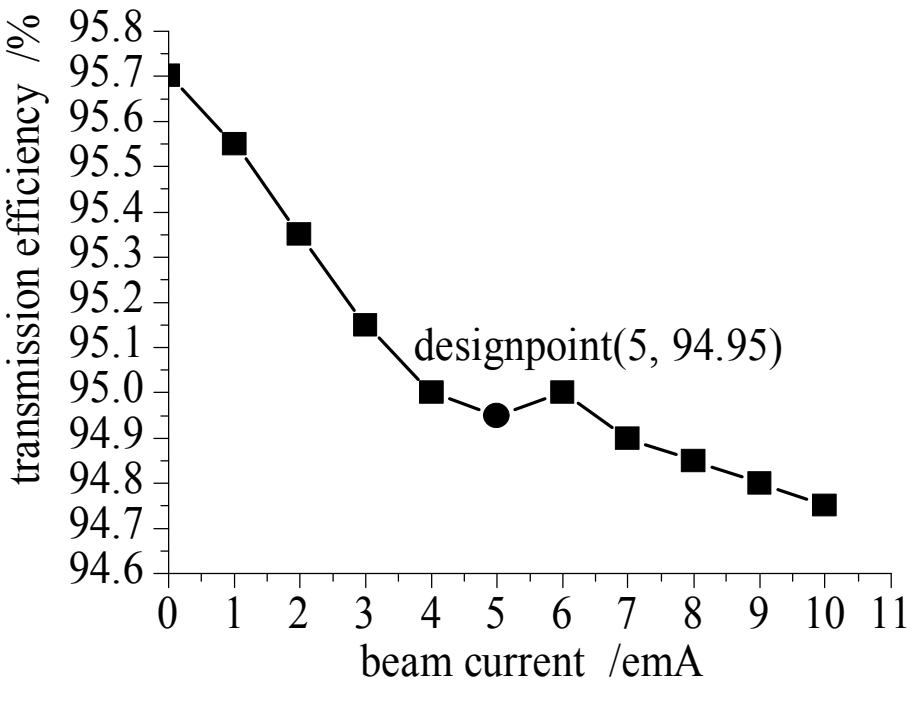

Fig 6 beam transmission efficiency

## 3 Discussions of the principal parameters

### 3.1 Acceleration gradient

Effective voltage of accelerating gap is shown in Figure 7. Because of the different RF ramping at both ends of the cavity, and RF coupling of triplet between neighboring two adjacent 0 degree KONUS sections, the effective accelerating voltage for these gaps are set as half of normal RF amplitude. The difference between the measured and designed effective gap voltage distributions will be tuned by the cavity tuning. In order to tune conveniently, an approximately constant maximum on-axis electric field along the whole structure are initially designed. As the increasing of gap length, the gap voltage distribution is ramped from about 260 kV at the low-energy end to about 393 kV at high energy end. Maximum value of axis field is shown in Figure 8. Because of CW operation, according to design experience of IH-DTL of GSI High Charge State Injector， the design value of the averaged acceleration gradient along cavity is about 2.48MV/m. Maximum value of axis field is 10.2MV/m, meanwhile Kilpatrick breakdown field is 10.56MV/m at 81.25MHz.

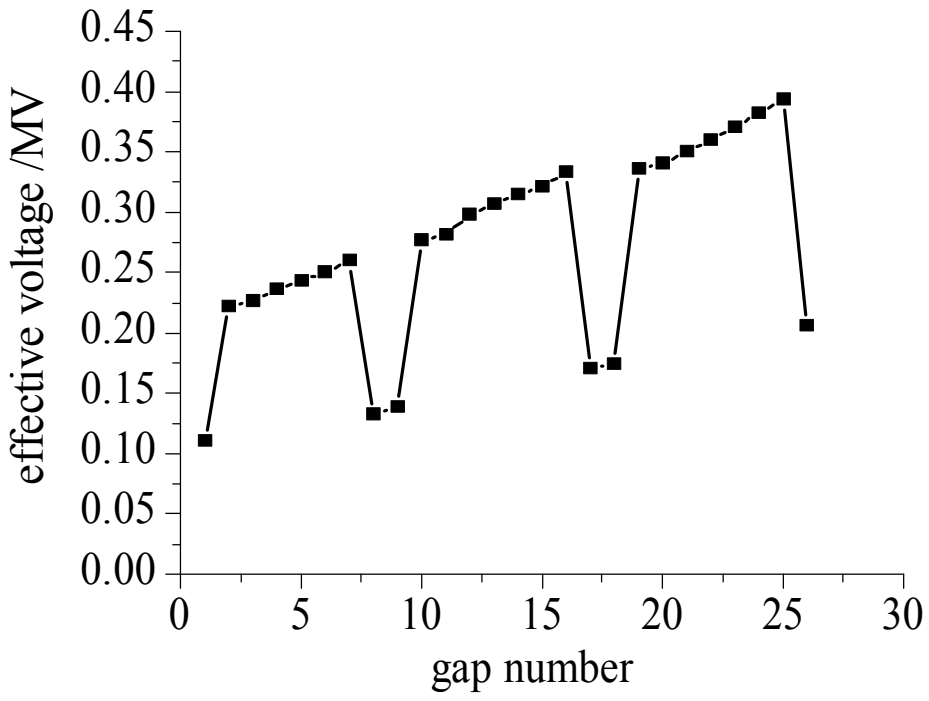

Fig 7 Effective voltage in accelerating gap

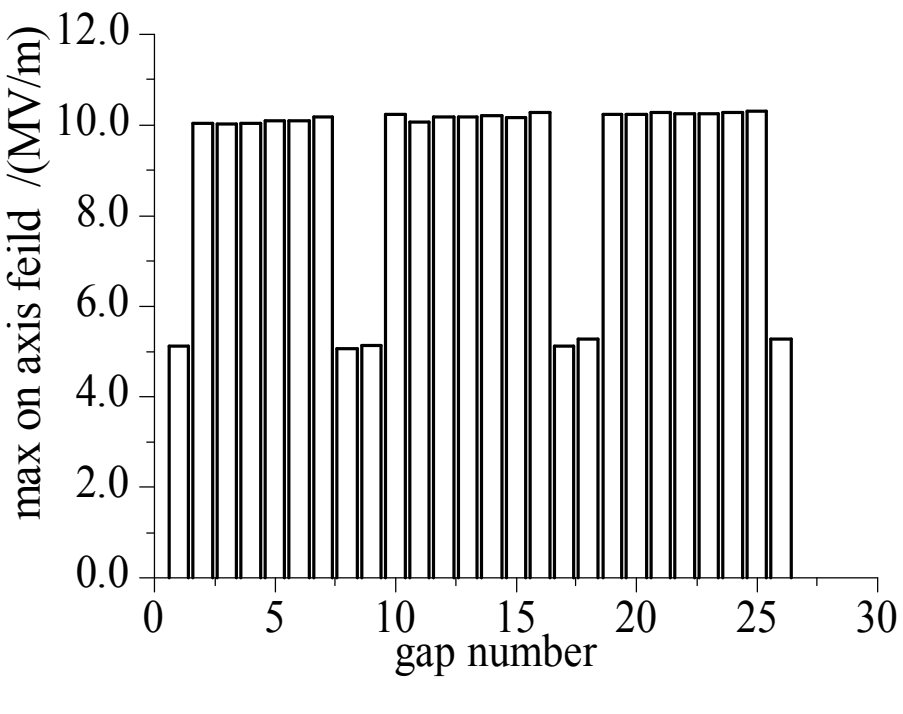

Fig 8 Maximum value of axis field

### 3.2 Quadruple field strengths

Parameters of magnetic quadruple triplet lens are listed in Table 2. Two quadruple triplets (QT1 and QT2) are placed in the IH-DTL to divide the IH-DTL into three accelerating sections. For the longitudinal motion the lens acts like a drift space and must be as short as possible. Shorter powerful quadruple triplet lenses are needed for sufficient transverse focusing and minimum longitudinal bunched phase expanding. The maximum quadruple field gradient is 98T/m， and the corresponding pole tip magnetic field is 1.176T.

Table 2. Quadruple lens parameters

| Triplet lenses | QT1 | QT2 | QT3 |
|---|---|---|---|
| Eff. Space (mm) | 20/20 | 20/20 | 20/20 |
| Eff. Pole length (mm) | 44/79/44 | 46/84/46 | 46/90/46 |
| Field gradient (T/m) | 96/ 98/96 | 88.5/92/88.5 | 93.5/90/93.5 |

### 3.3 H-type structure

IH-DTL has very high effective shunt impedance, which lead to shorter cavity length and less power consumption. CH-structure has an excellent mechanical strength and it is convenient for water cooling with respect to CW operation mode, however the dimension of CH-structure is too large at 81.25MHz. The cooling design for IH-DTL is the key point for the structure design. This will be simulated by the CST Studio in the near future.

## 4 Conclusions

A compact KONUS beam dynamics design for the low β IH-DTL of $^{238}U^{34+}$ beam is preliminary investigated with LORASR code. It will accelerate 5.0 emA $^{238}U^{34+}$ beams from 0.35 MeV/u to 1.30 MeV/u by an 81.25 MHz IH-DTL cavity. It achieves the transmission efficiency of 94.95% with the cavity length of 267.8 cm. The optimization has made emittance growth only about 8.0%. The designed average acceleration gradient is about 2.48MV/m.

The authors would like to thank Prof. U. Ratzinger at Frankfurt University and Prof. Li Zhihui at Sichuan University for their kind help and useful discussions. This work was supported by NSFC, grand No. 11079001，PKU and IMP.